\documentclass{article}
\usepackage{hyperref}
\usepackage[left=1.5in,top=1in,bottom=1in,right=1.5in]{geometry}

\begin{document}
\begin{center}
\huge{{\bf Gravity as Nonmetricity}}
\end{center}

\begin{center}
\Large{{\bf General Relativity in Metric-Affine Space (}{\bf {\it
L}}$_{n}${\bf ,}{\bf {\it g}}{\bf )}}
\end{center}

\begin{center}
\textbf{Alexander Poltorak}\\\small{General Patent Corporation, 75
Montebello Road, Suffern, NY 10901 USA.\\E-mail:
apoltorak@gpci.com.}
\end{center}

\begin{center}
\textbf{Abstract}
\end{center}

In this paper we propose a new geometric interpretation for
General Relativity (GR). It has always been presumed that the
gravitational field is described in GR by a Levi-Civita
connection. We suggest that this may not necessarily be the case.
We show that in the presence of an arbitrary affine connection,
the gravitational field is described as nonmetricity of the affine
connection. An affine connection can be interpreted as induced by
a frame of reference (FR), in which the gravitational field is
considered. This leads to some interesting observations, among
which: (a) gravity is a nonmetricity of space-time; (b) the affine
curvature of space-time induced in a noninertial FR contributes to
the stress-energy tensor of matter as an additional source of
gravity; and (c) the scalar curvature of the affine connection
plays the role of a ``cosmological constant''. It is interesting
to note that although the gravitational field equations are
identical to Einstein's equations of GR, this formulation leads to
a covariant tensor (instead of the pseudotensor) of
energy-momentum of the gravitational field and covariant
conservation laws. We further develop a geometric representation
of FR as a metric-affine space, with transition between FRs
represented as affine deformation of the connection. We show that
the affine connection of a NIFR has curvature and may have
torsion. We calculate the curvature for the uniformly accelerated
FR. Finally, we show that GR is inadequate to describe the
gravitational field in a NIFR. We propose a generalization of GR
that describes gravity as nonmetricity of the affine connection
induced in a FR. The field equations coincide with Einstein's
except that all partial derivatives of the metric are replaced by
covariant derivatives with respect to the affine connection. This
generalization contains GR as a special case of the inertial FR.
\medskip


\medskip
\bigskip

\Large{\textbf{ Introduction}} \normalsize
\medskip

In the Riemannian space $V_{4}$ of General Relativity (GR) two
principal geometric objects, metric \textbf{\emph{g}} and
connection $\Gamma$, are linked through the requirement of metric
homogeneity, i.e. the covariant derivative of metric vanishes
identically: $\nabla g=0$. This condition assures that the length
of a vector transported parallel in any direction remains
invariant. Since GR was first formulated, metric \textbf{\emph{g}}
and the Levi-Civita connection $\Gamma$ have been considered
respectively as a potential and strength of the gravitational
field. It is easy to see that the well-known difficulties, such as
non-covariance of the energy-momentum pseudo-tensor of the
gravitational field, that have plagued GR are directly related to
the choice of noncovariant connection $\Gamma$ (which is not a
tensor) as the strength of a gravitational field. We will endeavor
to demonstrate in the following that this need not be the case. In
part I we show that in the presence of an arbitrary affine
connection, the Einstein field equations lend themselves to a
novel geometrical interpretation wherein the affine deformation
tensor of the Levi-Civita connection plays the role of a
gravitational field. Furthermore, in the case of an affine
connection with vanishing torsion, the gravitational field becomes
the nonmetricity of spacetime. In this section we are not
concerned with the nature of this auxiliary affine connection and
can consider it as merely a convenient device. The fact that these
results hold true for any auxiliary affine connection suggests
that this geometric interpretation is merely a recasting of GR in
a new light, which does not change the field equations or any of
the predictions of the theory. The advantage of this geometric
interpretation of gravity as affine deformation or nonmetricity is
that it leads to a fully covariant theory with a true tensor for
energy momentum of the gravitational field.
\medskip

In part II we show that, guided by the ideas of geometrodynamics,
we are compelled to describe inertial forces, as well as gravity,
as affine deformation or nonmetricity.  In this regard, the
results obtained in Part I appear not at all surprising. In this
section we also offer a possible physical interpretation of an
auxiliary affine connection as induced by a chosen frame of
reference, in which the field is considered. This interpretation
of the auxiliary affine connection as no longer optional, but
rather as a required part of the description of physical reality,
makes the new geometric interpretation of GR offered herein so
much more compelling.  We describe the geometry in a noninertial
frame of reference and calculate the curvature in a uniformly
accelerating FR.
\medskip

In part III we consider gravity in a noninertial frame of
reference.  We demonstrate that GR is unsuitable for describing
the gravitational field in a noninertial FR.  We propose here a
generalization of GR wherein the gravitational field is a
nonmetricity of the affine connection induced in a chosen FR.  We
show here that the inertial forces play the role of a gauge field,
which must be turned on to compensate for the choice of a
noninertial FR.  This theory contains GR as a special case of the
gravitational field in an inertial frame of reference.

\bigskip

\section{General Relativity in a Metric-Affine Space (L$_{n}$,g)}

We start with Riemannian space $V_{4}$ -- the standard geometrical
setup of Einstein's GR comprising a differential manifold $M_{4}$
with metric \emph{\textbf{g}} and Levi-Civita connection $\Gamma$.
Let us introduce some arbitrary affine connection $\overline
\Gamma$ on the same manifold $M_{4}$. This auxiliary connection is
in no way linked with either the metric \emph{\textbf{g}} nor with
the Levi-Civita connection $\Gamma$. Such a geometric structure is
usually called a metric-affine space ($L_{4}$,$g)$. It is
important to bear in mind that at this point, the affine
connection $\overline \Gamma$ has no particular meaning and is
purely arbitrary.\footnote{ The use of an auxiliary geometric
device as an aid in the study of the subject at hand is not
unusual in geometry where it is common, for example, to study
$n$-dimensional manifolds as submerged in manifolds of higher
dimension. For the time being, we shall consider our auxiliary
affine connection as merely an aid in the study of the geometrical
characteristics of the gravitational field. The physical meaning
of this connection will be clarified at a later point. }
\medskip

As is well-known in differential geometry, affine connection
$\overline \Gamma$ on a differential manifold $M$ with the metric
\emph{\textbf{g}} can be always decomposed into the sum of the
Levi-Civita (metric) connection $\Gamma$, nonmetricity \textbf{S}
and torsion \textbf{Q}:

\begin{equation}
\label{eq1} \overline \Gamma   = \Gamma+S+Q.
\end{equation}

\noindent Note that connections $\overline \Gamma$ and $\Gamma$
are not tensors, while nonmetricity \textbf{S} and torsion
\textbf{Q} are tensors. In a local chart $x$ we can write the
expression (\ref{eq1}) in components:

\begin{equation}
\label{eq2} \overline \Gamma _{\mu \nu }^\lambda =\Gamma _{\mu \nu
}^\lambda +S_{\mu \nu }^\lambda +Q_{\mu \nu }^\lambda.
\end{equation}

The tensor of nonmetricity $S^{\lambda }_{\mu \nu }$ is symmetric
in its two lower indices while the torsion tensor $Q^{\lambda
}_{\mu \nu }$ is antisymmetric in the two lower indices. An affine
connection with vanishing nonmetricity, (\textbf{S}=0, $\overline
\Gamma$=$\Gamma$+\textbf{Q}), is called a Riemannian connection. A
Riemannian connection without torsion, (\textbf{Q}=0, $\overline
\Gamma$=$\Gamma$), is called a Levi-Civita connection. If the
affine connection has a vanishing torsion and, therefore, is
symmetric, we will denote it as $\bar {\Gamma }$. Although torsion
may play an important role both in field theory and in the
description of non-inertial frames of reference, for the sake of
simplicity, unless otherwise stated we shall assume vanishing
torsion, \textbf{Q}=0. Thus, all affine connections considered
herein are symmetric.
\medskip

Let $g_{\mu \nu }$ be a metric tensor; $\overline {\nabla }$ be a
covariant derivative with respect to the affine connection
$\overline \Gamma^{\lambda }_{\mu \nu }$. The nonmetricity tensor
$S^{\lambda }_{\mu \nu }$ can be expressed as

\begin{equation}
\label{eq3} g_{\tau \sigma } S_{\mu \nu }^\tau =\frac{1}{2}\left(
{\overline \nabla _\mu g_{\nu \sigma } +\overline \nabla _\nu
g_{\mu \sigma } -\overline \nabla _\sigma g_{\mu \nu } } \right)
\end{equation}

\noindent or

\begin{equation}
\label{eq4} g_{\tau \sigma } S_{\mu \nu }^\tau =\frac{1}{2}\left(
{\rho _{\mu \nu \sigma } +\rho _{\nu \mu \sigma } -\rho _{\sigma
\mu \nu } } \right),
\end{equation}

\noindent where

\begin{equation}
\label{eq5} \rho _{\mu \nu \sigma } =\overline \nabla _\mu g_{\nu
\sigma }
\end{equation}

\noindent is the metric inhomogeneity tensor.
\medskip

For any given Levi-Civita connection $\Gamma$ and affine
connection $\bar {\Gamma }$, there is a unique decomposition
\cite{Poltorak:1980} of the connection

\begin{equation}
\label{eq6}
\begin{array}{l}
 \Gamma =\bar {\Gamma }-{\rm {\bf S}}, \\
 \Gamma _{\mu \nu }^\lambda =\bar {\Gamma }_{\mu \nu }^\lambda -S_{\mu \nu
}^\lambda; \\
 \end{array}
\end{equation}

\noindent of the Riemann curvature tensor

\begin{equation}
\label{eq7}
\begin{array}{l}
 {\rm {\bf R=\bar {R}-\hat {R}}}, \\
 R_{\lambda \mu \nu }^\varepsilon =\bar {R}_{\lambda \mu \nu }^\varepsilon
-\hat {R}_{\lambda \mu \nu }^\varepsilon; \\
 \end{array}
\end{equation}

\noindent of the Ricci tensor

\begin{equation}
\label{eq8}
\begin{array}{l}
 {\rm {\bf Rc=\bar {R}c-\hat {R}c}}, \\
 R_{\mu \nu } =\bar {R}_{\mu \nu } -\hat {R}_{\mu \nu }; \\
 \end{array}
\end{equation}

\noindent of the scalar curvature

\begin{equation}
\label{eq9} R=\bar {R}-\hat {R};
\end{equation}

\noindent and of the Einstein tensor

\begin{equation}
\label{eq10}
\begin{array}{l}
 {\rm {\bf G=\bar {G}-\hat {G}}}, \\
 G_{\mu \nu } =\bar {G}_{\mu \nu } -\hat {G}_{\mu \nu }; \\
 \end{array}
\end{equation}

\noindent where {\bf R }($R^{\varepsilon }_{\lambda \mu \nu )}$,
{\bf Rc }($R_{\mu \nu )}$, $R$, and {\bf G} ($G_{\mu \nu })$ are
respectively the Riemann curvature tensor, Ricci tensor, scalar
curvature, and Einstein tensor, $G_{\mu \nu }\equiv  R_{\mu \nu
}$--$\raise.5ex\hbox{$\scriptstyle 1$}\kern-.1em/
\kern-.15em\lower.25ex\hbox{$\scriptstyle 2$} ${\it Rg}$_{\mu \nu
}$, of the Levi-Civita connection $\Gamma$; ${\rm {\bf \bar
{R}}}\left( {\bar {R}_{\lambda \mu \nu }^\varepsilon } \right)$,
${\rm {\bf \bar {R}c}}\left( {\bar {R}_{\mu \nu } } \right)$,
$\bar {R}$, and ${\rm {\bf \bar {G}}}\left( {\bar {G}_{\mu \nu } }
\right)$ are respectively the Riemann curvature tensor, Ricci
tensor, scalar curvature and Einstein tensor of the affine
connection $\bar {\Gamma }$; ${\rm {\bf \hat {R}}}\left( {\hat
{R}_{\lambda \mu \nu }^\varepsilon } \right)$, ${\rm {\bf \hat
{R}c}}\left( {\hat {R}_{\mu \nu } } \right)$, $\hat {R}$, and
${\rm {\bf \hat {G}}}\left( {\hat {G}_{\mu \nu } } \right)$ are
nonmetric components of, respectively, the Riemann curvature
tensor, Ricci tensor, scalar curvature and Einstein tensor of the
affine connection $\overline \Gamma$ defined as follows:

\begin{equation}
\label{eq11} \hat {R}_{\lambda \mu \nu }^\varepsilon =\overline
\nabla _\lambda S_{\mu \nu }^\varepsilon -\overline \nabla _\mu
S_{\lambda \nu }^\varepsilon +S_{\lambda \tau }^\varepsilon S_{\mu
\nu }^\tau -S_{\lambda \mu }^\varepsilon S_{\nu \tau }^\tau,
\end{equation}

\noindent wherein $\hat {R}_{\mu \nu } \equiv \hat {R}_{\lambda
\mu \nu }^\lambda $, $\hat {R}\equiv \hat {R}_{\mu \nu } g^{\mu
\nu }$and

\begin{equation}
\label{eq12} \hat {G}_{\mu \nu } \equiv \hat {R}_{\mu \nu }
-\frac{1}{2}g_{\mu \nu } \hat {R}.
\end{equation}
\bigskip

It may be useful to summarize these definitions in the following
table:
\begin{quote}
\begin{table}[htbp]
\begin{center}
\begin{tabular}{|p{75pt}|l|p{75pt}|p{75pt}|r|}
\hline Geometric Object& Levi-Civita Connection $\Gamma$
($\Gamma$$^{\lambda }_{\mu \nu })$& Affine Connection $\bar
{\Gamma }\left( {\bar {\Gamma }_{\mu \nu }^\lambda } \right)$&
Nonmetricity Tensor {\bf S} ($S^{\lambda }_{\mu \nu })$ \\
\hline Riemann Curvature Tensor& {\bf R} ($R^{\varepsilon
}_{\lambda \mu \nu })$& ${\rm {\bf \bar {R}}}\left( {\bar
{R}_{\lambda \mu \nu }^\varepsilon } \right)$&
${\rm {\bf \hat {R}}}\left( {\hat {R}_{\lambda \mu \nu }^\varepsilon } \right)$ \\
\hline Ricci Tensor& {\bf Rc} ($R_{\mu \nu })$& ${\rm {\bf \bar
{R}c}}\left( {\bar {R}_{\mu \nu } } \right)$&
${\rm {\bf \hat {R}c}}\left( {\hat {R}_{\mu \nu } } \right)$  \\
\hline Scalar Curvature& $R$& $\bar {R}$&
$\hat {R}$ \\
\hline Einstein Tensor& {\bf G} ($G_{\mu \nu })$& ${\rm {\bf \bar
{G}}}\left( {\bar {G}_{\mu \nu } } \right)$&
${\rm {\bf \hat {G}}}\left( {\hat {G}_{\mu \nu } } \right)$  \\
\hline
\end{tabular}
\label{tab1}
\end{center}
\end{table}
\end{quote}
\bigskip
\bigskip

Let us now consider Einstein's equation for the gravitational
field:

\begin{equation}
\label{eq13}
\begin{array}{l}
 {\rm {\bf G}}=8\pi {\rm {\bf T}}, \\
 G_{\mu \nu } =8\pi T_{\mu \nu }. \\
 \end{array}
\end{equation}

We use here geometrical units wherein the speed of light constant
and the gravitational constant of Newton are both set to unity. In
view of (\ref{eq10}) the expression (\ref{eq13}) can be recast as
\cite{Poltorak:1980}
\begin{equation}
\label{eq14}
\begin{array}{l}
 {\rm {\bf \hat {G}}}=8\pi {\rm {\bf \bar {T}}}, \\
 \hat {G}_{\mu \nu } =8\pi \bar {T}_{\mu \nu }, \\
 \end{array}
\end{equation}

\noindent where the modified stress-energy tensor T is defined as

\begin{equation}
\label{eq15}
\begin{array}{l}
 {\rm {\bf \bar {T}}}={\rm {\bf T}}-\frac{1}{8\pi }{\rm {\bf \bar {G}}}, \\
 \bar {T}_{\mu \nu } =T_{\mu \nu } -\frac{1}{8\pi }\bar {G}_{\mu \nu }. \\
 \end{array}
\end{equation}

It is easy to see that the tensor ${\rm {\bf \hat {G}}}$ has
vanishing covariant divergence with respect to the Levi-Civita
connection $\Gamma$: $\hat {G}_{\mu ;\nu }^\nu =0$, i.e. the
Bianchi identity holds true, and consequently the new
stress-energy tensor ${\rm {\bf \hat {T}}}$ satisfies the
conservation laws: $T_{\mu ;\nu }^\nu =0$. Thus, equations
(\ref{eq14})-(\ref{eq15}) can serve as the equations for the
gravitational field. These equations look very much like
Einstein's original field equation (\ref{eq13}). In fact, these
equations are equivalent to Einstein's equations, from which they
were derived. Since the unique decomposition of the Levi-Civita
connection and all curvature tensors into their respective affine
and nonmetric components (\ref{eq6})--(\ref{eq10}) hold true for
any affine connection $\bar {\Gamma }$, which is unrelated to the
gravitational field or its source, these equations are identically
equivalent to Einstein's standard equation (\ref{eq13}) and we are
still on the firm ground of classical General Relativity. And yet,
these equations in the form (\ref{eq14})-(\ref{eq15}) present
quite a different geometrical picture. The left side of equation
(\ref{eq14}) describes the gravitational field as the nonmetricity
of the chosen affine connection, which in turn contributes the
stress-energy tensor T as an additional source of the
gravitational field.
\medskip

Let us emphasize that unlike metric-affine theories of
gravitation\cite{Hehl:1995}, which consider the metric and
connection (and, sometimes, the coframe) to be independent field
potentials, this reformulation of GR still considers only the
metric to be the gravitational potential. The affine connection is
unrelated to the gravitational field and is not a dynamic
variable. It is defined ad hoc so that its curvature tensors may
be computed outside of the field equations. It is most convenient
to choose a connection with zero curvature, which is just as
suitable for our purposes, although any other connection may be
used.
\medskip

What is the physical meaning of the affine connection $\overline
\Gamma$? As we shall see in the next section, this connection may
represent the geometry of the frame of reference (FR), in which
the gravitational field is considered. Obviously, the geometry of
the FR, i.e. the affine connection $\bar {\Gamma }$, does not
depend on the stress-energy tensor {\bf T} and does not represent
the gravitational field. Therefore, in \cite{Poltorak:1980}, we
moved the affine Einstein tensor ${\rm {\bf \bar {G}}}$ to the
right side of the equation, which also reflects the fact that the
inertial forces generated in a noninertial frame of reference have
energy and, therefore, contribute to the stress-energy tensor of
matter as an additional source of the gravitational field.
Consequently, the gravity is now described by the nonmetricity of
the spacetime. Indeed, the nonmetricity tensor {\bf S} ($S^{
\lambda }_{\mu \nu })$ describes the strength of the gravitational
field in equation (\ref{eq14}).
\medskip

As we take a fresh look at our field equations\cite{Poltorak:1980}
(\ref{eq14})-(\ref{eq15}), we notice that the affine part of the
Einstein tensor $\bar {G}_{\mu \nu } \equiv \bar {R}_{\mu \nu }
-\frac{1}{2}\bar {R}g_{\mu \nu } $ contains field potentials --
metric tensor $g_{\mu \nu }$ -- and, therefore, can hardly be
justified as the field source as part of the stress-energy tensor.
The only part that is independent of the gravitational field is
the affine Ricci tensor $\bar {R}_{\mu \nu } $, which is
calculated based on the affine connection $\bar {\Gamma }$ set
{\it a priori}. Consequently, we shall modify our field equation
as follows:

\begin{equation}
\label{eq16} \hat {G}_{\mu \nu } -\frac{1}{2}\bar {R}g_{\mu \nu }
=8\pi \hat {T}_{\mu \nu },
\end{equation}

\noindent where the modified stress-energy tensor $\hat {T}_{\mu
\nu } $ is defined as

\begin{equation}
\label{eq17} \hat {T}_{\mu \nu } =T_{\mu \nu } -\frac{1}{8\pi
}\bar {R}_{\mu \nu }.
\end{equation}

It is easy to recognize in the field equation (\ref{eq16}) the
structure of Einstein's equation with a cosmological constant
wherein the affine scalar curvature $\bar {R}$ plays the role of
the cosmological constant $\Lambda $, although our ``cosmological
constant'' is not necessarily constant. This curious similarity
notwithstanding, the field equations (\ref{eq16})-(\ref{eq17}) are
identically equivalent to the classical Einstein field equations
(\ref{eq13}) \emph{without} the cosmological constant.
\medskip

The fact that the field equations (\ref{eq16})-(\ref{eq17}) are
equivalent to the standard equations of GR guarantees that there
will be no tests that can distinguish between the two
interpretations. It is then legitimate to ask, what are the
advantages of this new geometrical interpretation of GR? The
answer lies in the mathematical rigor and physical meaningfulness
of the theory.
\smallskip

As has been known for a long time, GR suffers from certain
difficulties related to the lack of general covariance of the
theory. To wit, the energy-momentum pseudo-tensor of the
gravitational field is not a covariant object, which leads to a
lack of local conservation laws for the gravitational field -- an
unacceptable situation in our view. A popular attempt to explain
away this difficulty by the principle of equivalence appears to be
misguided.  It fact, the principle of equivalence itself is not
well defined in GR. This principle establishing local equivalence
of the gravitational field and inertial forces arising in a
noninertial FR, first of all, requires a good definition of the
frame of reference, which, unfortunately, is all too often
confused with a coordinate system. From here it is deduced that
since gravity vanishes in a free-falling FR, there is nothing
wrong with gravity vanishing in Riemannian coordinates, in which
the Levi-Civita connection is zero.  The fallacy of this argument
is rooted in equating the Riemannian coordinate system with a
free-falling FR.  In this metric-affine reformulation of GR, the
energy-momentum of the gravitational field is described by a
covariant tensor.\cite{Poltorak:1980, Poltorak:1983} The global
conservation laws also exist in this
framework.\cite{Poltorak:1980, Poltorak:1983} Indeed, the Bianchi
identity requires that

\begin{equation}
\begin{array}{l}
 \nabla T = 0, \\
 {\rm T}_{\nu ;\mu }^\mu   = 0{\rm  }{\rm .} \\
 \end{array}
\end{equation}

These conservation laws can be rewritten in our formalism as

\begin{equation}
\begin{array}{l}
 \overline \nabla  \left( {T + t} \right) = 0, \\
 {\rm T}_{\nu |\mu }^\mu   + {\rm t}_{\nu |\mu }^\mu   = 0{\rm  ,} \\
 \end{array}
\end{equation}

\noindent where the semicolon denotes a covariant derivative with
respect to Levi-Civita connection $\Gamma $, a vertical line
denotes a covariant derivative with respect to affine connection
${\bar \Gamma }$, and t ($t_\nu ^\mu$) is now a true tensor
obeying covariant conservation laws.
\medskip

Let us consider, for example, a trivial affine connection $\bar
{\Gamma }$ with zero curvature. In this case, the Ricci tensor
$\bar {R}_{\mu \nu } $ and the scalar curvature $\bar {R}$ of this
affine connection $\bar {\Gamma }$ vanish. The nonmetric Einstein
tensor $\hat {G}_{\mu \nu } $ differs from the regular Einstein
tensor $G_{\mu \nu }$ only in that all partial derivatives of the
metric are replaced with the covariant derivatives with respect to
the affine connection $\bar {\Gamma }$: $\partial _\mu   \to
\delta _\mu ^\lambda  \bar \nabla _\nu   = \delta _\mu ^\lambda
\partial _\nu   - \bar \Gamma _{\mu \nu }^\lambda  $.  We see that the flat affine connection $\bar \Gamma $ plays here the role of a gauge field compensating for the
arbitrary coordinate transformation and assuring general
covariance of the theory.
\medskip

To summarize, the mere existence of an auxiliary affine connection
$\bar {\Gamma }$ allowed us to recast the Einstein field equations
(without really changing them) in a form that suggests a novel
geometrical interpretation of gravity as the nonmetricity of
spacetime. This reformulation of General Relativity allows for
ridding the theory of its difficulties related to noncovariance.

\section{Geometrodynamics in a Frame of Reference}

In the previous section we made a suggestion that the affine
connection of the metric-affine space ($L_{4}$,$g)$ may represent
a frame of reference. In this section we will justify this
hypothesis and show how the affine connection is determined in a
chosen frame of reference. We will consider the concept of
geometrodynamics as the guiding principle in describing any
``universal''\cite{Reichenbach:1958} force such as gravitational
or inertial. Thus, the objective of this analysis is to find an
appropriate geometric description of the noninertial frames of
reference (NIFR) and the transformation laws between different
frames of reference.
\medskip

Einstein's GR, despite its claim to be the general theory of
relativity, does not even define frames of reference. The
principle of relativity is replaced by the principle of general
covariance, confusing reference frames with coordinate systems,
which play little role in the geometry of spacetime. This position
is untenable because coordinate systems have no physical meaning
whatsoever, while the frame of reference is a fundamental physical
concept. A particular choice of a FR affects the physical laws
therein.
\medskip

As has been pointed out by Kretschman\cite{Kretschman:1917},
Fock\cite{Fock:1967}, Wigner\cite{Wigner:1963},
Rodichev\cite{Rodichev:1974}, Mitzkevich\cite{Mitzkevich:1971} and
a few other authors, the coordinate system is merely a way to
number points or label events of spacetime.\cite{Misner:1973}
Therefore, the general covariance principle is seen as devoid of
physical meaning and a mere triviality.\cite{Ohanian:1976} We can
well formulate both the geometry of spacetime and the physics in a
given spacetime in the coordinate-free language of contemporary
mathematics. It is for the purpose of illustrating this very point
that we provided duplicative coordinate free representation for
most of the above equations and geometrical objects.
\medskip

There have been a number of attempts to describe frames of
reference as chronometric invariants, monads or {\bf $\tau
$}-fields and tetrads. We shall explore here another approach that
stems directly from the very notion of geometrodynamics.

\subsection{Special Relativity in an arbitrary coordinate system}
\label{subsec:mylabel1}

As is well-known, according to the Special Theory of Relativity,
the geometry in inertial frames of reference is the
pseudo-Euclidean geometry of Minkowski four-dimensional spacetime.
\medskip

In special relativity it is accepted that (a) IFRs are represented
by Lorentz Coordinate systems and that Lorentz transformations,
which are representations of the Lorentz group of rotation in the
Minkowski spacetime, describe the transition from one IFR to
another. Generally speaking, coordinate systems, being merely a
scheme of numbering points on a manifold (in our case, the events
of the spacetime continuum), are devoid of any physical meaning
and do not describe any reference frames, inertial or noninertial.
Different IFRs are represented by different Minkowski spaces and
the transition from one IFR to another is a diffeomorphism $\Re
:M^1 \to M^2 $.
\medskip

A Minkowski space is a four-dimensional differential manifold
whose points are spacetime events, with a Minkowski metric {\bf
{\it $\eta $}} defined on the manifold. Assuming that different
observers experience the same events albeit from different vantage
points\footnote{We are disregarding here the fact that an event
visible to one observer may not be visible to another observer if
this event lies outside his event horizon.}, we can assume that
all events constitute a single manifold and different inertial
observers correspond to different Minkowski metrics defined over
the same differential manifold. Thus, the transition from one IFR
to another in geometric terms amounts to the transformation {\bf
{\it $\eta $}}$^{1}\to $ {\bf {\it $\eta $}}$^{2}$. More
precisely, since, generally, the curvature is determined by the
connection rather than by the metric, although Minkowski space has
no curvature, nevertheless, it more correct to say that the
transition from one IFR to another in geometric terms amounts to
the transformation $\Gamma ^1 \to \Gamma ^2 $\footnote{Needless to
say, in Minkowski space the connection happens to be metric
compatible, i.e. a flat Levi-Civita connection.}. Due to the fact
that Minkowski space, as a pseudo-Euclidian space, is flat,
different Minkowski spaces, i.e. different Minkowski metrics on
the manifold, are essentially identical up to a general coordinate
transformation. Such coordinate transformations do not affect the
metric, which is invariant, but they do affect the connection,
which is not a covariant object. It is important to remember that
an observer in an IFR is free to choose any coordinate system,
which does not necessarily have to be a Lorentz (pseudo-Cartesian)
coordinate system. Of course, in a flat space, such as Minkowski,
it is always possible (and preferable) to select a global
orthogonal pseudo-Cartesian coordinate system such as Lorentz, in
which all connection coefficients vanish globally, as is
frequently done; but this is an option, not a requirement.
\medskip

These facts explain why associating IFR with Lorentz coordinate
systems in Special Relativity, just as the use of Galilean
coordinates to describe IFR in Newtonian mechanics (in a flat 3+1
Euclidian space), is acceptable for all practical purposes (albeit
conceptually misleading) and does not lead to any contradictions.
This situation changes radically as we attempt to describe
noninertial frames of reference. A failure to recognize that a
coordinate transformation, no matter how complex, can never
describe a transition from an IFR to a NIFR or from one NIFR to
another NIFR, has led to much confusion.

\subsection{Geometrodynamics in a Noninertial frame of reference}

Let us consider an observer traveling in a spacecraft. Suppose
that from the point of view of another inertial observer, the
spacecraft, which we will consider to be the reference body of the
NIFR associated with the observer traveling therein, is
accelerating with a constant acceleration {\bf {\it \textbf{a}}}
along a straight line in a 3-D space, which translates into a
hyperbolic motion in the Minkowski 4-D spacetime of the inertial
observer. If the observer in the spacecraft observes a few test
particles freely moving inside the craft, she will notice that
they all move with acceleration \textbf{--}{\bf {\it \textbf{a}}}
in the direction opposite of the direction of the spacecraft (as
indicated by the accelerometers). If the test particles made of
different material all accelerate uniformly, this suggests to the
observer that either these particles move under the influence of a
universal force (in Reichenbach's terminology) or that the
spacetime is non-Euclidean. Thus, considering the motion of the
test particles, the observer in the NIFR of the spacecraft has two
choices:

\begin{enumerate}
\item to postulate the Minkowski spacetime and assume existence of a universal force, which acts upon these test particles and causes them all to accelerate with acceleration --{\bf {\it \textbf{a}}}, or
\item to rule out any universal forces and to admit that the geometry inside the spacecraft is {\it not} Euclidian (or rather not pseudo-Euclidian, i.e. not Minkowski), i.e. the spacetime is not flat.
\end{enumerate}

As Poincar\'e pointed out very early (and Reichenbach stressed
later), the reality is the sum total of physics and geometry:

\begin{center}EMPIRICAL REALITY = PHYSICS + GEOMETRY
\end{center}

Although one is free to choose where the separation line between
physics and geometry lies and, therefore, each of the two choices
above are legitimate, Poincar\'e and Reichenbach advocated the
second choice whereby all universal forces are eliminated.  This
is the principle of geometrodynamics.
\medskip

Pursuant to the second choice, as dictated by geometrodynamics,
the geometry of NIFR is non-Euclidean. The question remains,
however, how to determine precisely the geometry within a NIFR.
\medskip

To do that, let us first consider the movement of a spacecraft in
an inertial frame of reference.
\medskip

Let $M$ be Minkowski spacetime with metric {\bf {\it $\eta $}} and
connection $\bar {\Gamma }$, which in a local chart $x$ have
respective components of {\it $\eta $}$_{\mu \nu }$ and $\bar
{\Gamma }_{\mu \nu }^\lambda $.  All freely moving test particles
in the Minkowski space of an IFR obey the following equation:\

\begin{equation}
\label{eq18} \bar {\nabla }_X X=0,
\end{equation}

\noindent where $\bar {\nabla }$ is a covariant derivative with
respect to the connection $\bar {\Gamma }$. In the local chart $x$
the equation (\ref{eq18}) takes the form:

\begin{equation}
\label{eq19} \frac{d^2x^\lambda }{d\tau ^2}+\bar {\Gamma }_{\mu
\nu }^\lambda \frac{dx^\mu }{d\tau }\frac{dx^\nu }{d\tau }=0,
\end{equation}

\noindent where {\it $\tau $} is a smooth affine parameter along
the worldline, which can be taken to represent the proper time of
this test particle.
\medskip

Expression (\ref{eq19}) is the equation for a geodesic line in an
arbitrary coordinate system $x$. Similarly, the equation for a
reference body (in our example, the spacecraft) accelerating with
acceleration {\bf {\it a}} takes the form of

\begin{equation}
\label{eq20} \bar {\nabla }_X X=a,
\end{equation}

\noindent or

\begin{equation}
\label{eq21} \frac{d^2x^\lambda }{d\tau ^2}+\bar {\Gamma }_{\mu
\nu }^\lambda \frac{dx^\mu }{d\tau }\frac{dx^\nu }{d\tau
}=a^\lambda.
\end{equation}

Let us now consider the movement of the same test particles from
within the spacecraft, i.e., from a NIFR. All of the test
particles inside the spacecraft are accelerating with respect to
an observer inside the craft with the same acceleration {\bf {\it
a}} but in the opposite direction:

\begin{equation}
\bar \nabla _X X =  - a,
\end{equation}

\noindent or

\begin{equation}
\label{eq22} \frac{d^2x^\lambda }{d\tau ^2}+\bar {\Gamma }_{\mu
\nu }^\lambda \frac{dx^\mu }{d\tau }\frac{dx^\nu }{d\tau
}=-a^\lambda.
\end{equation}

If we do not insist on maintaining flat Minkowski geometry and do
not wish to admit the existence of universal forces causing
acceleration --{\bf {\it \textbf{a}}}, i.e. we choose a
geometrodynamical representation of reality, we have to assume
that these test particles move along the geodesic lines of a
non-Euclidian space
--- a space of affine connection $\Gamma $:

\begin{equation}
\label{eq23} \nabla _X X=0,
\end{equation}

\noindent or

\begin{equation}
\label{eq24} \frac{d^2x^\lambda }{d\tau ^2}+\Gamma _{\mu \nu
}^\lambda \frac{dx^\mu }{d\tau }\frac{dx^\nu }{d\tau }=0,
\end{equation}

\noindent where $\nabla $ is a covariant derivative with respect
to the affine connection $\Gamma $ and $\Gamma ^{\lambda }_{\mu
\nu }$ are the components of the connection $\Gamma $. Eliminating
the force acting on a test particle by describing its motion as a
free fall along geodesics in a non-Euclidean space is the essence
of geometrodynamics, which aims to describe the field of force as
a manifestation of non-Euclidean geometry of spacetime.
\medskip

Note that the equations (\ref{eq22}) and (\ref{eq24}) describe the
same trajectory of the same test particle. (Since the affine
parameter {\it $\tau $} is related only to the curve representing
the trajectory of a test particle, which is a geometrical
invariant, we are justified in using the same affine parameter for
(\ref{eq22}) and (\ref{eq24}) describing the same curve.) Hence,
deducting (\ref{eq22}) from (\ref{eq24}) we obtain

\begin{equation}
\label{eq25} T_{\mu \nu }^\lambda \frac{dx^\mu }{d\tau
}\frac{dx^\nu }{d\tau }=a^\lambda,
\end{equation}

\noindent where

\begin{equation}
\label{eq26} T_{\mu \nu }^\lambda =\Gamma _{\mu \nu }^\lambda
-\bar {\Gamma }_{\mu \nu }^\lambda
\end{equation}

\noindent is called the tensor of affine deformation. It is easy
to see that the Minkowski metric {\bf {\it $\eta $}} is
inhomogeneous with respect to the affine connection $\Gamma $,
i.e. its covariant derivative with respect to this connection does
not vanish: $\nabla \eta  \ne 0$. Generally, according to
(\ref{eq1}), the affine deformation $T^{\lambda }_{\mu \nu }$ is
comprised of the symmetric tensor of nonmetricity $S^{\lambda
}_{\mu \nu }$ and anti-symmetric torsion tensor $Q^{\lambda }_{\mu
\nu }$:

\begin{equation}
\label{eq27} T_{\mu \nu }^\lambda =S_{\mu \nu }^\lambda +Q_{\mu
\nu }^\lambda.
\end{equation}

As is known, torsion does not affect geodesics, i.e. two affine
connections different only by torsion have the same geodesics.
Thus, for non-rotating FRs, we can disregard torsion and assume
that affine connection $\Gamma ^{\lambda }_{\mu \nu }$ is
symmetric in its two lower indices:

\begin{equation}
\label{eq28} \Gamma _{\mu \nu }^\lambda =\Gamma _{\nu \mu
}^\lambda.
\end{equation}

\noindent and, therefore, the tensor of affine deformation is
equal to the tensor of nonmetricity:

\begin{equation}
\label{eq29} T_{\mu \nu }^\lambda =S_{\mu \nu }^\lambda.
\end{equation}

The tensor of nonmetricity can be expressed through the covariant
derivatives of metric as follows:

\begin{equation}
\label{eq30} \eta _{\tau \sigma } S_{\mu \nu }^\tau
=\frac{1}{2}\left( {\nabla _\mu \eta _{\nu \sigma } +\nabla _\nu
\eta _{\mu \sigma } -\nabla _\sigma \eta _{\mu \nu } } \right).
\end{equation}

Furthermore, it is interesting to note that any geodesic
transformation of the affine connection, i.e. a transformation of
the type:

\begin{equation}
\label{eq31} \tilde {\Gamma }_{\mu \nu }^\lambda =\Gamma _{\nu \mu
}^\lambda +\frac{1}{2}\left( {p_\mu \delta _\nu ^\lambda +p_\nu
\delta _\mu ^\lambda } \right),
\end{equation}

\noindent where $p_{\mu }$ is an arbitrary covariant vector, does
not affect the geodesics. Consequently, the geometry of a NIFR
based on the worldline geodesics is defined only up to an
arbitrary torsion and an arbitrary geodesic transformation of the
type (\ref{eq31}).

\subsection{Geodesics or Autoparallel lines?}
\label{subsec:mylabel2}

Since it turns out that the space in a NIFR is a space of affine
connection with nonmetricity (and, possibly, torsion), we have to
retrace our steps and take a closer look at the step leading to
equations (\ref{eq23}) and (\ref{eq24}). We called the
trajectories of the test particles geodesic lines. There is a
certain inconsistency in this terminology that may lead to
confusion. In geometry, the ``straightest'' line defined by the
parallel transport of a tangent vector, i.e. by the affine
connection, is always called a geodesic line. The shortest line
or, more generally, the line of a stationary length, which
requires a metric, is called the extremal path of a certain
functional. In Riemannian space, where the Levi-Civita connection
is metrically compatible, the straightest lines coincide with the
lines of stationary length and both are called geodesics. In a
space of affine connection these two types of curves no longer
coincide. In physics literature, it has become accepted to call
the ``shortest'' line, or rather, the line of extremal length,
{\it geodesic}, and the straightest line, {\it auto-parallel}. The
line with the extremal proper length is derived through a
variational principle of Euler-Lagrange:

\begin{equation}
\label{eq32} \delta s=\int\limits_A^B {ds} =\int\limits_A^B
{\left( {-g_{\mu \nu } dx^\mu dx^\nu } \right)^{1/2}} =0
\end{equation}

\noindent or

\begin{equation}
\label{eq33} \frac{\delta I}{\delta x^\sigma
}=\frac{1}{2}\int\limits_A^B {g_{\mu \nu } \frac{dx^\mu }{d\lambda
}\frac{dx^\nu }{d\lambda }d\lambda } =0
\end{equation}

The ``straightest'' line, on the other hand, is an expression of
parallel transport defined by connection, merely requiring that a
vector is transported parallel to itself along the line $s$, which
is called an autoparallel line. Which line, the ``straightest'' or
the ``shortest,'' are we to take as an expression of the
trajectory of a test particle?
\medskip

In our view, the ``shortest'' line is not a local concept. It
requires a line between two points A and B to have the shortest
(or extremal) proper length (or time for a timelike worldline). A
test particle in any given point on its trajectory ``knows''
nothing about the length of the curve between the point where the
particle is and some other point where it {\it is not}.
Consequently, it is illogical to assume that the freely moving
test particle will follow the ``shortest'' line. Such an
assumption would imply that the particle in a given point on the
worldline somehow is aware of the global properties of this
worldline extending into the future.
\medskip

On the other hand, the ``straightest'' or autoparallel line is a
local concept. In any given point on the line, the connection and
curvature of the line are defined in that point. This curvature
can be minimized (or extremized) in that point. Thus the
straightest line is a logical choice for a test particle to
follow. Henceforth, we shall continue to use only the
``straightest,'' i.e. auto-parallel, lines to describe
trajectories of free test particles, but will retain for them the
term {\it geodesics} as it is accepted in the literature on
differential geometry.

\subsection{Geometry in Noninertial Frames of Reference}

Let us now proceed to calculate the curvature of a NIFR. To
achieve this goal we need to find the solution to equation
(\ref{eq25}), which we will rewrite here in a slightly different
form:

\begin{equation}
\label{eq34} T_{\mu \nu }^\lambda u^\mu u^\nu =a^\lambda,
\end{equation}

\noindent where the local velocity of the test particle is denoted
as $u^{\mu }$={\it dx}$^{\mu }$/{\it dt}. Suppose $C_{\mu \nu }$
is a covariant tensor of the second rank. Let us consider
projection of this tensor on the two velocity vectors $u^{\mu }$
and $u^{\nu }$:

\begin{equation}
\label{eq35} C_{\mu \nu } u^\mu u^\nu =c^2,
\end{equation}

\noindent where $c^{2}$ is an invariant scalar, which we for
simplicity shall consider a constant. Then a solution of the
equation for affine deformation (\ref{eq34}) takes the form:

\begin{equation}
\label{eq36} T_{\mu \nu }^\lambda =\frac{1}{c^2}a^\lambda C_{\mu
\nu. }
\end{equation}

Whenever an affine connection $\overline \Gamma$ is transformed
into a connection $\Gamma $ by affine deformation {\bf
T}\footnote{ This (1,2) tensor of affine deformation T (${\rm
T}_{\mu \nu }^\lambda $) should not be confused with the (0,2)
stress energy tensor T (${\rm T}_{\mu \nu }^{} $) used in the
field equations (13).}: $\bar {\Gamma }$= $\Gamma$ + {\bf T}, or
in components

\begin{equation}
\label{eq37} \Gamma _{\mu \nu }^\lambda =\bar {\Gamma }_{\nu \mu
}^\lambda +{\rm T}_{\nu \mu }^\lambda,
\end{equation}

\noindent the Riemannian curvature tensor \textbf{R }undergoes the
following transformation:

\begin{equation}
\label{eq38} R_{\lambda \mu \nu }^\sigma   = \bar R_{\lambda \mu
\nu }^\sigma   + \bar \nabla _\lambda  T_{\mu \nu }^\sigma   -
\bar \nabla _\mu  T_{\lambda \nu }^\sigma   + T_{\lambda \rho
}^\sigma  T_{\mu \nu }^\rho   - T_{\mu \rho }^\sigma  T_{\lambda
\nu }^\rho   + 2Q_{\lambda \mu }^\rho  T_{\rho \nu }^\sigma,
\end{equation}

\noindent where $R^{\sigma }_{\lambda \mu \nu }$ is the Riemannian
curvature tensor of the second affine connection $\Gamma$, $\bar
{R}_{\lambda \mu \nu }^\sigma $is the Riemannian curvature tensor
of the first connection $\bar {\Gamma }$ (in our case, this is the
Levi-Civita connection of the Minkowski space in an IFR), $\bar
{\nabla }$ is the covariant derivative with respect to the first
connection $\bar {\Gamma }$, {\bf T} ($T^{\lambda }_{\mu \nu })$
is the affine deformation and {\bf Q} ($Q^{\lambda }_{\mu \nu })$
is the torsion of the second connection $\Gamma$.
\medskip

Note that the Levi-Civita connection of the Minkowski space in IFR
is flat -- hence its curvature tensor is zero: $\bar {R}_{\lambda
\mu \nu }^\sigma =0$. Furthermore, since the torsion leaves
geodesics invariant, we will for now disregard it: $Q^{\lambda
}_{\mu \nu }$=0. Now expression (\ref{eq38}) takes a simpler form
of

\begin{equation}
\label{eq39} R_{\lambda \mu \nu }^\sigma =\bar {\nabla }_\lambda
T_{\mu \nu }^\sigma -\bar {\nabla }_\mu T_{\lambda \nu }^\sigma
+T_{\lambda \rho }^\sigma T_{\mu \nu }^\rho -T_{\mu \rho }^\sigma
T_{\lambda \nu }^\rho
\end{equation}

\noindent or

\begin{equation}
\label{eq40} R_{\lambda \mu \nu }^\sigma =\bar {\nabla }_{[\lambda
} T_{\mu ]\nu }^\sigma +T_{\rho [\lambda }^\sigma T_{\mu ]\nu
}^\rho.
\end{equation}
\medskip

Substituting in (\ref{eq39}) the value derived for the affine
deformation from (\ref{eq36}), we get

\begin{equation}
\label{eq41} R_{\lambda \mu \nu }^\sigma =\frac{1}{c^2}\left[
{\bar {\nabla }_\lambda \left( {a^\sigma C_{\mu \nu } }
\right)-\bar {\nabla }_\mu \left( {a^\sigma C_{\lambda \nu } }
\right)} \right]+\frac{1}{c^4}a^\sigma a^\rho \left( {C_{\lambda
\rho } C_{\mu \nu } -C_{\mu \rho } C_{\lambda \nu } } \right)
\end{equation}

\noindent or

\begin{equation}
\label{eq42} R_{\lambda \mu \nu }^\sigma =\frac{1}{c^2}\left(
{C_{\mu \nu } \bar {\nabla }_\lambda a^\sigma -C_{\lambda \nu }
\bar {\nabla }_\mu a^\sigma +a^\sigma \bar {\nabla }_{[\lambda }
C_{\mu ]\nu } } \right)+\frac{1}{c^4}a^\sigma a^\rho \left(
{C_{\lambda \rho } C_{\mu \nu } -C_{\mu \rho } C_{\lambda \nu } }
\right).
\end{equation}

The Ricci curvature tensor defined as $R_{\mu \nu }= R^{\lambda
}_{\lambda \mu \nu }$ takes the form

\begin{equation}
\label{eq43} R_{\lambda \mu \nu }^\lambda =\frac{1}{c^2}\left[
{\bar {\nabla }_\lambda \left( {a^\lambda C_{\mu \nu } }
\right)-\bar {\nabla }_\mu \left( {a^\lambda C_{\lambda \nu } }
\right)} \right]+\frac{1}{c^4}a^\lambda a^\rho \left( {C_{\lambda
\rho } C_{\mu \nu } -C_{\mu \rho } C_{\lambda \nu } } \right)
\end{equation}

\noindent or

\begin{equation}
\label{eq44} R_{\mu \nu } =\frac{1}{c^2}\left( {C_{\mu \nu } \bar
{\nabla }_\lambda a^\lambda -C_{\lambda \nu } \bar {\nabla }_\mu
a^\lambda +a^\lambda \bar {\nabla }_{[\lambda } C_{\mu ]\nu } }
\right)+\frac{1}{c^4}a^\lambda a^\rho \left( {C_{\lambda \rho }
C_{\mu \nu } -C_{\mu \rho } C_{\lambda \nu } } \right).
\end{equation}

We can also calculate the scalar curvature by contracting the
Ricci tensor $R_{\mu \nu }$ with the Minkowski metric {\it $\eta
$}$^{\mu \nu }$:

\begin{equation}
\label{eq45} R = \frac{1}{{c^2 }}\left( {C\bar \nabla _\lambda
a^\lambda   - C_\lambda ^\mu  \bar \nabla _\mu  a^\lambda   + \eta
^{\mu \nu } a^\lambda  \left[ {\bar \nabla _\lambda  C_\mu ^\mu -
\bar \nabla _\mu  C_\lambda ^\mu  } \right]} \right) +
\frac{1}{{c^4 }}a^\lambda  a^\rho  \left( {C_{\lambda \rho } C -
C_{\mu \rho } C_\lambda ^\mu  } \right)
\end{equation}

\noindent where C={\it  $\eta $}$^{\mu \nu }C_{\mu \nu .}$
\medskip

Needless to say, the four equations (37) are not enough to
uniquely define the tensor of affine deformation ${\rm T}_{\mu \nu
}^\lambda  $, which has 40 components.  We need an additional
assumption to determine the connection.  Let us now make some
assumptions about the tensor $C_{\mu \nu .}$ The simplest and the
most important covariant tensor of the second rank that exists in
our geometry is the Minkowski metric tensor {\it $\eta $}$_{\mu
\nu }$. It is also symmetric, as is $C_{\mu \nu }$, and appears to
be the most natural candidate for the role of $C_{\mu \nu }$.
Consequently, we are going to assume that

\begin{equation}
\label{eq46} \mbox{C}_{\mu \nu } =\eta _{\mu \nu }.
\end{equation}

\noindent The expression

\begin{equation}
T_{\mu \nu }^\lambda   = \frac{1}{{c^2 }}a^\lambda  \eta _{\mu \nu
}
\end{equation}

\noindent is certainly a solution of equation (\ref{eq34}) and
seems to be the most meaningful physically. This assumption allows
us to rewrite the expressions for the curvature tensors as
follows:

\begin{equation}
\label{eq47} R_{\lambda \mu \nu }^\sigma =\frac{1}{c^2}\left(
{\eta _{\mu \nu } \bar {\nabla }_\lambda a^\sigma -\eta _{\lambda
\nu } \bar {\nabla }_\mu a^\sigma } \right)+\frac{1}{c^4}a^\sigma
\left( {a_\lambda \eta _{\mu \nu } -a_\mu \eta _{\lambda \nu } }
\right),
\end{equation}

\begin{equation}
\label{eq48} R_{\mu \nu } =\frac{1}{c^2}\left( {\eta _{\mu \nu }
\bar {\nabla }_\lambda a^\lambda -\bar {\nabla }_\mu a_\nu }
\right)+\frac{1}{c^4}\left( {a^2\eta _{\mu \nu } -a_\mu a_\nu }
\right),
\end{equation}

\begin{equation}
\label{eq49} R=\frac{3}{c^2}\left( {\bar {\nabla }_\lambda
a^\lambda +\frac{a^2}{c^2}} \right).
\end{equation}

Let us suppose now that the NIFR is uniformly accelerating, i.e.,
the acceleration \emph{\textbf{a}} is constant. That assumption
allows us to further simplify the above expressions:

\begin{equation}
\label{eq50} R_{\lambda \mu \nu }^\sigma =\frac{1}{c^4}a^\sigma
\left( {a_\lambda \eta _{\mu \nu } -a_\mu \eta _{\lambda \nu } }
\right),
\end{equation}

\begin{equation}
\label{eq51} R_{\mu \nu } =\frac{1}{c^4}\left( {a^2\eta _{\mu \nu
} -a_\mu a_\nu } \right),
\end{equation}

\begin{equation}
\label{eq52} R=\frac{3a^2}{c^4}.
\end{equation}

Although in geometric units, an assumption $\mbox{C}_{\mu \nu }
=\eta _{\mu \nu }$ necessarily leads c=1, because the square of
the proper velocity vector is a unity: $u^2  = u^\mu  u_\mu   = 1$
, it is informative to consider the expression (56) in real
physical units.  It is easy to see that the constant $c$ has
physical units of velocity [m/sec] and, in fact, coincides with
the speed of light. We see that although acceleration causes
curvature of spacetime, this curvature is remarkably small -- on
the order of 1/$c^{4}$.

\section{General Relativity in a Noninertial frame of reference}
\label{sec:general}

The discussion in Part I, in which we analyzed Einstein's field
equation in the presence of an arbitrary affine connection, has
revealed that the gravitational field ought to be described as
nonmetricity of spacetime. More specifically, the field equations
would look like Einstein's standard equations wherein the partial
derivatives of the metric tensor are replaced by covariant
derivatives with respect to the affine connection. However, even
in their modified form (14)-(15), Einstein's equations of GR are
hardly suited to describe the gravitational field in a noninertial
frame of reference. Indeed, we started our analysis with the
assumptions that the ``ultimate'' geometry is that of Riemann and
that the test particles move along the geodesics of the
Levi-Civita connection, which we merely decomposed into its affine
and nonmetric components. The reality may not be that simple, and
there is no reason to believe that the geometry in a noninertial
frame of reference is described by a Levi-Civita connection.  In
fact, from the above analysis, we can easily see that it is not
the case.
\medskip

Let us consider the evolution of the geometry one step at a time.
Let us start with the inertial frame of reference in the absence
of the gravitational field. According to the special theory of
relativity, the geometry in such a case is that of a Minkowski
space with a flat metric and a compatible Levi-Civita connection
$\mathop \Gamma \limits^0 $.
\medskip

As the next step, let us consider a noninertial frame of
reference. According to our analysis in Part II, the
transformation from an IFR to a NIFR will subject our Levi-Civita
connection to an affine deformation (\ref{eq26}) resulting in a
metric-affine space ($L_{4}$,$g)$ with the Minkowski metric {\bf
{\it $\eta $}} and the independent affine connection $\Gamma$
defined as

\begin{equation}
\label{eq53} \mathop \Gamma \limits^1 =\mathop \Gamma \limits^0
+\mathop T\limits^1,
\end{equation}

\noindent where the affine deformation $\mathop T\limits^1 $ is
defined by (\ref{eq37}).
\medskip

Finally, let us introduce a gravitational field into this NIFR. As
we concluded before, the gravity should be described as
nonmetricity. Thus, our affine connection of NIFR $\mathop \Gamma
\limits^1 $ undergoes another affine deformation

\begin{equation}
\label{eq54} \mathop \Gamma \limits^2 =\mathop \Gamma \limits^1
+\mathop {\rm {\bf T}}\limits^2,
\end{equation}

\noindent where the affine deformation $\mathop {\rm {\bf
T}}\limits^2 $ is none other than a nonmetricity tensor \textbf{S}
defined in (\ref{eq3}), which is the strength of the gravitational
field. Let us rewrite this expression in a more convenient format

\begin{equation}
\label{eq55} \Gamma =\bar {\Gamma }+{\rm {\bf S}},
\end{equation}

\noindent where we have adapted the following notations: $\bar
\Gamma = \mathop \Gamma \limits^1 $, $\Gamma =\mathop \Gamma
\limits^2 $ and ${\rm {\bf S}}=\mathop {\rm {\bf T}}\limits^2 $.
Consequently, the curvature tensors can be similarly decomposed as

\begin{equation}
\label{eq56} {\rm {\bf R}}={\rm {\bf \bar {R}}}+{\rm {\bf \hat
{R}}}
\end{equation}

\begin{equation}
\label{eq57} {\rm {\bf Rc}}={\rm {\bf \bar {R}c}}+{\rm {\bf \hat
{R}c}}
\end{equation}

\begin{equation}
\label{eq58} R=\bar {R}+\hat {R}
\end{equation}

\begin{equation}
\label{eq59} {\rm {\bf G}}={\rm {\bf \bar {G}}}+{\rm {\bf \hat
{G}}}
\end{equation}

Einstein's equations neatly follow from the Bianchi identity,
which requires that the covariant divergence of the Einstein
tensor vanishes.  The proportionality of the Einstein tensor
\textbf{G} to the stress-energy tensor \textbf{T} assures,
therefore, the conservation laws. We require that this
proportionality of the Einstein tensor \textbf{G} to the
stress-energy tensor \textbf{T} hold true in a metric affine space
of a noninertial frame of reference, i.e. that the Einstein tensor
\textbf{G} of the affine connection $\Gamma$ is still proportional
to the stress-energy tensor \textbf{T}, which may now include the
contribution of the inertial forces that have energy and,
therefore, are an additional source of the gravitational field. It
is only naturally to assume that such a contribution, i.e. the
stress-energy tensor of the inertial forces, is equal to the
Einstein tensor of the affine connection $\bar {\Gamma }$: ${\rm
{\bf T}}_{inert} ={\rm {\bf \bar {G}}}$. Thus we have:

\begin{equation}
\label{eq60} {\rm {\bf \bar {G}}}+{\rm {\bf \hat {G}}}=8\pi {\rm
{\bf T}}+{\rm {\bf \bar {G}}}.
\end{equation}

\noindent Canceling the affine Einstein tensor on both sides of
the equation, we arrive at the equations for the gravitational
field in a noninertial frame of reference:

\begin{equation}
\label{eq61}
\begin{array}{l}
 {\rm {\bf \hat {G}}}=8\pi {\rm {\bf T,}} \\
 \hat {G}_{\mu \nu } =8\pi T_{\mu \nu. } \\
 \end{array}
\end{equation}

As is easily seen, the equations (\ref{eq61}) are Einstein's
equations (\ref{eq13}) in which all partial derivatives of the
metric {\bf {\it g}} are replaced by the covariant derivatives
with respect to the affine connection $\bar {\Gamma }$.

\begin{equation}
\partial
_\mu   \to \delta _\mu ^\lambda  \bar \nabla _\nu   = \delta _\mu
^\lambda  \partial _\nu   - \bar \Gamma _{\mu \nu }^\lambda
\end{equation}
\medskip

These equations are no longer equivalent to Einstein's equations
and represent a generalization of Einstein's theory for a
gravitational field in a noninertial frame of reference. In the
simplest case of the inertial frame of reference, the equations
(\ref{eq61}) are equivalent to Einstein's equation (\ref{eq13}).
Therefore, this generalization contains the classical GR in a
special case of the gravitational field in an inertial frame of
reference.

\section{Conclusion}
\label{sec:conclusion}

Analyzing GR in the presence of an arbitrary affine connection, we
have determined that Einstein's equations describe the
gravitational field as nonmetricity of an auxiliary affine
connection. Since this result holds true for any affine
connection, including a flat connection, we concluded that GR in
fact describes gravity as nonmetricity of spacetime.
\medskip

Analysis of the geometry in a noninertial frame of reference
revealed that this is a metric-affine geometry and that the
transformation between frames of reference is represented as an
affine deformation.
\medskip

It was concluded that GR, which demands Riemannian geometry, is
inadequate to describe a gravitational field in a non-inertial
frame of reference. A simple generalization of General Relativity
we proposed has the field equations that revert to Einstein's
equations as a special case of an inertial frame of reference.

\end{document}